\documentclass[sigconf]{acmart}

\usepackage{booktabs} 
\usepackage{textcomp}

\setcopyright{none}

\acmConference[]{SenSys '17}{November 6--8, 2017}{Delft, Netherlands}  
\acmYear{2017}
\copyrightyear{2017}
\acmPrice{0.00}

\begin{document}
\title{Zebra-RFO - A Spectrum Repository for the Masses}

\author{Andr\'es Arcia-Moret}
\affiliation{%
  \institution{Computer Laboratory\\
University of Cambridge}
}
\email{andres.arcia@cl.cam.ac.uk}

\renewcommand{\shortauthors}{A. Arcia-Moret}

\begin{abstract}
TV White Spaces has recently been in the interest of the networking community as an alternative to alleviate the spectrum crunch, incentivizing the need to understand the dynamics of congestion of the occupied spectrum and the quantification of the free spectrum. In this respect, many regulatory organizations may provide references for the legal allocation of the spectrum and therefore allow primary and secondary users to plan their deployments. In this article, we present the motivations and challenges to collect spectrum measurements as a global challenge. We discuss a prototype to massively collect spectrum footprints at low-cost to make it available to communities.  
\end{abstract}

%
%



\maketitle

\section{Introduction}
The convenience of wireless networks regarding mobility and ease of deployment has made them hugely popular. These networks convey data in the order of couple of exa-bytes per month and, by 2019, this number is expected to grow at least in one order of magnitude\footnote{Cisco Visual Networking Index: Global Mobile Data Traffic Forecast Update 2014--2019 White Paper.}. A natural consequence of this tendency is a congested wireless spectrum creating the so-called \textit{spectrum crunch}. 

In emerging countries, regulators use manual and static databases to keep track of incumbents and secondary users of the wireless spectrum; thus, leading to inaccurate and cumbersome information of the spectrum occupancy.  Moreover, there are intermittent legal users (e.g., UHF microphones), unaccounted legal users, and rogue users that make use of the spectrum with no control, being potential interferers. \textit{This is a clear opportunity for regulators and local authorities to promote regionalized and distributed (i.e., localized) repositories} for keeping track of the used and unused frequencies, boosting a more efficient use of the spectrum considering a regulatory framework \cite{arcia-moret-gaia}. Moreover, providing communities with low-cost commercial off-the-shelf technology for assessing the spectrum usage is a promising way of tackling the spectrum crunch \cite{xia}.

Recent wireless technologies such as TV White Spaces (TVWS) can be deployed if there is enough information about unoccupied portions of the spectrum. TVWS networks can be deployed in rural and remote areas more easily because this technology is meant to overcome long distances. However, a successful deployment depends on the availability of spectrum, but accounting the free spectrum has always been an expensive task \cite{arcia-moret-gaia}. The cost of spectrum analyzers is in the order of tens of thousand of dollars and the processing of information generated by these devices is not oriented towards assessing the available frequencies of interest (e.g., white spaces) within a geographical region.

Lightweight, containerizable and distributed repositories relying on a central authority (for regulation purposes) will allow people and governments to cooperate, paving the way to alternative wireless network deployments; bringing Internet connectivity for the masses \cite{arcia-moret-gaia}. Flexible TVWS repository placement will allow a timely report of the spectrum occupation and will enable better content delivery and adequate support for the local production of content and services. 

We are developing an ecosystem of open and low-cost devices and services to ascertain the current occupation of UHF and ISM band; allowing capturing and processing spectrum occupation in under-served areas.  Our aim is to incentivize communities to be aware of the local occupation the spectrum of interest. Zebra-RFO (Radio Frequency Observer)\footnote{http://www.zebra-rfo.org} is a (containerizable) web service with collaboration capabilities akin to social networks, able to organize long measurement campaigns allowing the visualisation of spectrum occupation. Zebra-RFO also offers the possibility of editing measurement campaigns in order to isolate different areas of interest (i.e., rural, urban, suburban), and also conveniently represent the rough occupation of large portions of the spectrum in UHF band and ISM band, both of high interest in the process of bringing the next billion people on-line\footnote{https://trac.tools.ietf.org/html/rfc7962}.

\subsection{Challenges}

Many challenges have to be overcome when designing a light-weight and containerizable repository while considering different specific regulations imposed by governments: 

 \textbf{Massification of the spectrum repositories}. With the aid of low-cost spectrum monitoring devices and Community Network (CN) infrastructure, containerized spectrum repositories can be conveniently distributed wihtin the CN. A flexible naming system, as proposed in Information Centric approaches, will allow users a transparent localized access to the closest repository; thereby having a timely response to occupation. The distributed nature of the spectrum data collection incentivises multiple regional clients to coexist. Eventually, the information has to be consensual at the central authority and then pushed back to the regional repositories \cite{arcia-moret-gaia}. \textit{Zebra-RFO relies on light-weight consensus protocols to have a uniform vision of the spectrum} \cite{heidi}.

 \textbf{Providing an API to enable spectrum governance}. Our system aims to provide a consistent and public view of the regulator vision of the spectrum which may be subject to non compatible rules for assigning and accessing spectrum occupation among conflicting parties. It is well known that in emerging regions there is a departure from international recommendations (e.g., ITU) for spectrum allocation\footnote{http://goo.gl/Znmjna}. \textit{Zebra-RFO will provide an API allowing the community negotiation of conflicting spectrum occupation}.

 \textbf{Crowdsourcing spectrum}. Promoting accessibility and understanding of occupation information by lay people so that spectrum collection is not of exclusive use of (business-oriented) primary users. Although orthogonal to the storage and retrieval problem, \textit{championing spectrum collection is a major concern for the success of any high-scale crowdsourcing initiative} \cite{licia}.

\section{Low-Cost Spectrum Collection}
\label{devices}
As conventional spectrum analysers are expensive, difficult to transport, and they lack an appropriate interface to collect continuous spectrum activity; there is a need for low-cost and long-term monitoring in developing regions. Several such devices for the sub 1-GHz band have recently been developed: Whisppy monitor using a Raspberry Pi to interface RFExplorer, ASCII32 monitor using SI4313/Arduino, or RFTrack. These solutions do not exceed the 400~US\$ price. However, they pose a heterogeneous vision of the spectrum usage, i.e., they use a different antenna, radio and sampling rates, in different geographical positions for capturing and uploading spectrum activity. Moreover, programmers tend to customize the capturing of the spectrum activity into different file formats, thus the need for a unifying approach provided by Zebra-RFO. 

\section{containerized repository}

\begin{figure}
\vspace{0.5cm}
\includegraphics[scale=0.19]{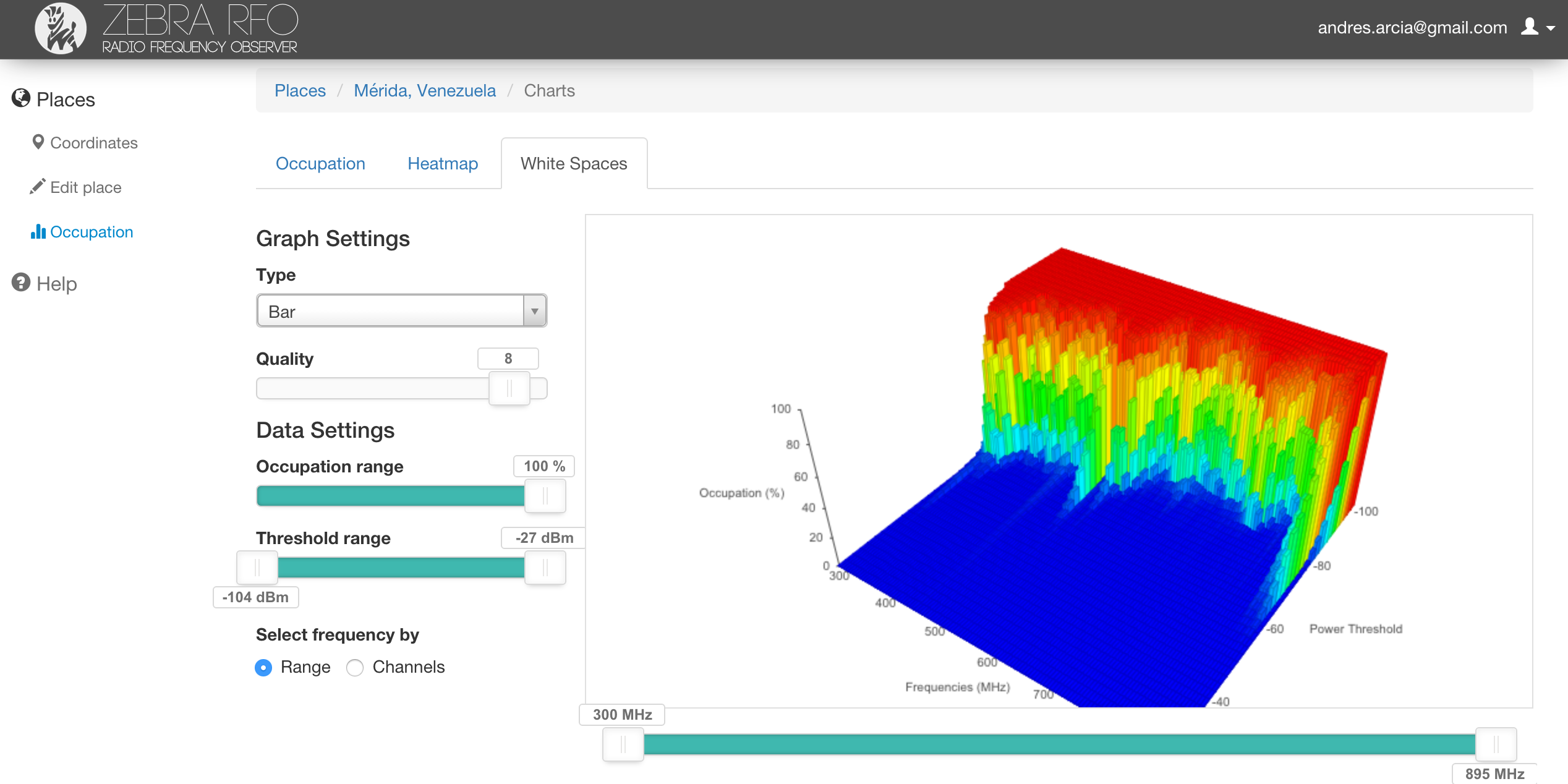}
\caption{White Spaces report on Zebra-RFO}
\label{fig:zebra-img}
\end{figure}

Fig.~\ref{fig:zebra-img} shows the operating prototype of a central repository called Zebra-RFO, an open initiative to collect spectrum footprints and a social platform to incentivize a crowd-sharing approach for collection. Our system provides data organization and visualization capabilities allowing later post-processing. Zebra-RFO offers capabilities such as convenient edition of the geo-tagged journeys to get rid of potential biases induced by the speed of mobile collectors. Moreover, data editing allows filtering areas of interest with the aid of a visual interface, i.e., well-defined urban areas, or rural areas in which networks based on TVWS could be deployed.

\section{Zebra Radio Frequency Observatory}

Zebra-RFO platform is currently running at the University of Cambridge cloud. There you can create an account and observe data collected from 4 different continents; specifically, the sample database contains open data\footnote{e.g., part of it comes from https://goo.gl/Jq5m8E} coming from 10 different countries, exploring rural and urban areas. In the sample account you can edit different data collections coming from the different devices mentioned in Section \ref{devices}, and will be able to observe the effect of variations of the detection threshold for spectra occupation. Through different graphical representations and simple GUI, users are able not only to transform the geo-tagged collections but also to have an overview of the spectrum occupation in developing regions.

\section{Acknowledgments}
The research leading to these results has received partial funding from the British Engineering and Physical Science Research Council (EPSRC) project: Networks as a Service (EP/K031724/2). Zebra-RFO has been co-developed with Freddy Rond\'on. Special thanks to Ermanno Pietrosemoli and Marco Zennaro for their valuable comments.

%
\bibliographystyle{ACM-Reference-Format}

%
%


\end{document}